\documentclass[twocolumn,showpacs]{revtex4}
\usepackage{amssymb}
\usepackage{makeidx}
\usepackage{amsmath}
\usepackage{graphicx}
\usepackage{dcolumn}
\usepackage{bm}

\begin{document}

\title{Harmonics generation and intense terahertz radiation from polar
molecules at multiphoton resonant excitation in laser fields}
\author{H. K. Avetissian}
\email{avetissian@ysu.am}
\author{B. R. Avchyan}
\author{G. F. Mkrtchian}
\affiliation{Centre of Strong Field Physics, Yerevan State University, 1 A. Manukian,
Yerevan 0025, Armenia }
\date{\today }

\begin{abstract}
The coherent radiation spectrum of two-level polar molecules with UV
transition is studied at the multiphoton resonant excitation by a moderately
strong laser field. The spectrum corresponding to harmonic generation and
low-frequency radiation is investigated both analytically and numerically.
Specifically, a mechanism for generation of intense smoothly tunable
terahertz radiation by two-level molecular configurations (with dynamic
Stark shifts) possessing permanent dipole moments, is considered.
\end{abstract}

\maketitle



\section{ Introduction}

The interaction of strong (moderately) laser radiation with the matter
produces a number of significant processes that include the harmonic
generation. The latter is of particular interest because of importance and
potential applications of tunable, coherent, intense high-frequency
radiation. Generation of high-order laser harmonics is a well-approbated
process in the systems composed by atoms or small molecules and modeled most
successfully by a "three-step recollision" physical picture \cite{HHG1,HHG3}%
. The three-step theoretical models for high harmonic generation does not
include any participation of excited bound states in the emission of high
harmonics and is essentially valid for harmonics with energy well above the
ionization energy \cite{HHG1}.

On the other hand, various studies have shown that for the certain systems
such as quantum wells and large molecules \cite{QuantWells,LargeMol},
bound-bound transitions are more important for harmonic generation than the
coupling between the ground and the continuum state \cite{BoundCont}. This
mechanism of harmonic generation without ionization can not provide
high-frequency radiation, e.g. x-ray, but can be more efficient for
generation of moderately high harmonics, in particular, for UV/VUV
applications \cite{AAM}. For description of strong light scattering process
via bound-bound transitions, resonant interaction is of interest. Apart from
its pure theoretical interest as a simple model, the resonant interaction
regime may allow to increase considerably the efficiency of frequency
conversion.

Harmonic generation by a resonantly driven two-level atom has been studied
in \cite{KrMi}. However, a two-level atomic system meets difficulties since
the required for efficient multiphoton resonant excitation laser fields are
so strong that many atomic levels and continuum will play important role 
\cite{Duvall}. So, two-level atomic system is not a good quantitative model
for description of atomic response in case of multiphoton-resonant
excitation.

Nevertheless, in some atomic and especially molecular systems one can avoid
the difficulties with efficient multiphoton excitation based on the
two-level model.\ As it has been shown \cite{PDM,AM,Gib,AM2,AM3,ABM}, the
multiphoton resonant excitation of a quantum system subjected to a strong
laser field is effective when the quantum system has permanent dipole
moments (PDM) in the stationary states. Otherwise, the energies of the
excited states of a three-level atom should be close enough to each other
and the transition dipole moment between these states must be nonzero.
Furthermore, these systems have an advantage, which allow to generate
radiation with frequency much lower than laser frequency \cite{Kib, THz}.
The effects of diagonal dipole elements on harmonic generation at the
one-photon resonance has been considered in \cite{Calderon}\textit{. }%
However, for efficient generation of moderately high harmonics by optical
pulses one should consider systems with the bound-bound UV/VUV transitions
and, therefore, the multiphoton resonant interaction regime is of special
interest. The coherent scattering spectrum of a two-level system possessing
PDM at multiphoton resonant excitation has been considered in Ref. \cite{GO}%
. However, the presented in this paper analytical results for radiation
spectrum have only qualitative character, in the meantime, the numerical
calculations have been performed for such strong fields which break the
condition of multiphoton resonant excitation (see below Eq. (\ref{applic}))
and also could lead to detrimental ionization in real atomic/molecular
systems.

In the present paper, coherent light scattering by two-level polar molecules
at multiphoton resonant excitation in the field of a moderately strong laser
radiation is investigated. It is shown that the study of this process is
important for efficient population transfer, generation of moderately high
harmonics, as well as for generation of low-frequency radiation,
specifically, for realization of intense smoothly tunable terahertz
radiation sources. A simple analytic expression for the time-dependent mean
dipole moment, taking into account the dynamic Stark shifts, is obtained. In
particular, based on this expression results, concerning the main spectral
characteristics of considering process, are in good agreement with results
of performed numerical calculations. The effect of compensation of dynamic
Stark shift on the radiation spectrum is considered. We compare the spectra
of two-level systems with and without PDM and show the advantages of polar
systems for efficient generation of desirable radiation. Special attention
is paid to generation of coherent low-frequency radiation in such systems,
particularly, smoothly tunable, intense terahertz radiation\textrm{.}

It should be noted that the obtained results may be interesting for
different type coherent light sources, such as polar gases, certain organic
crystals or quantum dots and etc. \cite{P1,P2,P3,Kib,P4}. Besides, all
presented results can be scaled to other systems and diverse domains of the
electromagnetic spectrum.

The paper is organized as follows: in Sec. II we present the analytical
model and derive the coherent contribution to the harmonic and low-frequency
spectra. In Sec. III we present some results of numerical calculations of
the considered issue without multiphoton resonant approximation and compare
the obtained spectra with analytical results. Finally, conclusions are given
in Sec. IV.

\section{THEORY}

\subsection{Basic Model}

We consider a two-level quantum system possessing PDM driven by a
time-dependent, intense laser field. The Hamiltonian describing the
interaction of the laser field with the considered quantum system is given
within semiclassical dipole approximation by 
\begin{equation*}
\hat{H}=\left( \varepsilon _{1}+V_{11}\right) |1\rangle \langle 1|+\left(
\varepsilon _{2}+V_{22}\right) |2\rangle \langle 2|
\end{equation*}%
\begin{equation}
+\left( V_{12}|1\rangle \langle 2|+\mathrm{h.c.}\right) .  \label{Ham}
\end{equation}%
Here, $\varepsilon _{1}$ and $\varepsilon _{2}$ are the energies of the
stationary states $\left\vert 1\right\rangle $ and $\left\vert
2\right\rangle $ of unperturbed molecule, $\varepsilon _{2}>\varepsilon _{1}$%
. In Eq. (\ref{Ham})%
\begin{equation}
V_{\eta \nu }=-d_{\eta \nu }E_{0}\cos \omega _{0}t,  \label{Interact}
\end{equation}%
is the interaction part of Hamiltonian with real matrix element of the
electric dipole moment projection $d_{\eta \nu }=\mathbf{\hat{e}}\cdot 
\mathbf{d}_{\eta \nu }$, and pump wave field is taken to be linearly
polarized, with unit polarization vector $\mathbf{\hat{e}}$, slowly varying
amplitude $E_{0}$, and carrier frequency $\omega _{0}$. The diagonal terms
in (\ref{Interact}) describe the interaction due to the PDM and are crucial
for effective multiphoton coupling.

The technique which we employ in our calculations has been described earlier
in \cite{AAM}. First we obtain dynamic wave function of the system in the
pump laser field, i.e., the solution of the time-dependent Schr\"{o}dinger
equation with Hamiltonian (\ref{Ham}). From these solution we deduce formula
for expectation value of the time-dependent dipole-moment operator. Note
that dipole-moment form of radiation spectrum gives similar results with
dipole-acceleration and dipole-velocity forms in case of long pulses \cite%
{TC-B}.

We consider Schr\"{o}dinger equation: 
\begin{equation}
i\frac{\partial \left\vert \Psi (t)\right\rangle }{\partial t}=\hat{H}%
\left\vert \Psi (t)\right\rangle ,  \label{Schrod_eq}
\end{equation}%
with Hamiltonian (\ref{Ham}) at multiphoton resonant excitation regime and
assume that $\left\vert \delta _{n}\right\vert \ll \omega _{0}$, where $%
\delta _{n}$ is the $n$-photon resonance detuning, given by the relation:%
\begin{equation}
\delta _{n}=\varepsilon _{1}-\varepsilon _{2}+n\omega _{0}.  \label{detuning}
\end{equation}%
Here and below, unless stated otherwise, we employ atomic units ($\hbar
=e=m_{e}=1$). Our method of solving (\ref{Schrod_eq}) has been described in
detail in \cite{AM} and will not be repeated here. Under the generalized
rotating wave approximation, the time-dependent wave function can be
expanded as: 
\begin{equation*}
\left\vert \Psi (t)\right\rangle =e^{-i\varepsilon _{1}t}\left\{ \left[ 
\overline{a}_{1}(t)+\alpha _{1}(t)\right] \left\vert 1\right\rangle +\left[ 
\overline{a}_{2}(t)+\alpha _{2}(t)\right] \right.
\end{equation*}%
\begin{equation}
\left. \times \left[ \exp [-i(n\omega _{0}t+\int_{0}^{t}(V_{22}-V_{11})dt)]%
\right] \left\vert 2\right\rangle \right\} ,  \label{wf}
\end{equation}%
where $\overline{a}_{i}(t)$ are the time-averaged probability amplitudes and 
$\alpha _{i}(t)$ are rapidly changing functions over pump wave period. The
time-averaged amplitudes $\overline{a}_{i}(t)$ are:%
\begin{equation}
\overline{a}_{i}=\sum\limits_{j=1}^{2}C_{ij}\exp (i\lambda _{j}t),
\label{diff_sol}
\end{equation}%
where $C_{ij}$ are the constants of integration determined by the initial
conditions, and the factors $\lambda _{j}$ are the solutions of the
second-order characteristic equation:%
\begin{equation}
\left\Vert 
\begin{array}{cc}
\Delta _{n}-\lambda & -F_{n} \\ 
-F_{n} & -\Delta _{n}-\delta _{n}-\lambda%
\end{array}%
\right\Vert =0,  \label{matrix}
\end{equation}%
with the function:%
\begin{equation}
F_{n}=\frac{d_{12}}{d_{p}}n\omega _{0}J_{n}\left( Z\right) ,  \label{avr}
\end{equation}%
and detuning:%
\begin{equation}
\Delta _{n}=\omega _{0}\left( \frac{d_{12}}{d_{p}}\right)
^{2}\sum\limits_{k\neq n}\frac{k^{2}J_{k}^{2}\left( Z\right) }{k-n}.
\label{stark}
\end{equation}%
Here $d_{p}=d_{22}-d_{11}$ (let $d_{p}>0$) is the difference of PDM in two
stationary states -in the excited and ground states. The argument of the
ordinary Bessel function $J_{n}\left( Z\right) $ is the difference of dipole
interaction energies in units of the pump wave photon energy: $%
Z=d_{p}E_{0}/\omega _{0}$. The terms $F_{n}$ and $\Delta _{n}$ describe the
resonant coupling and dynamic Stark shift at $n$-photon resonance,
respectively. In deriving these equations we have applied well-known
expansion of exponent through Bessel functions with real arguments \cite%
{Watson}:%
\begin{equation}
e^{iZ\sin \alpha }=\sum\limits_{s=-\infty }^{\infty }J_{s}\left( Z\right)
e^{is\alpha }.  \label{bess}
\end{equation}%
\qquad Assuming smooth turn-on of the pump wave, the relation between the
rapidly and slowly oscillating parts of the probability amplitudes can be
written as:%
\begin{equation}
\alpha _{1}(t)=\overline{a}_{2}(t)\frac{d_{12}}{d_{p}}\sum\limits_{k\neq n}%
\frac{kJ_{k}\left( Z\right) e^{i(k-n)\omega _{0}t}}{k-n},  \label{rap1}
\end{equation}%
\begin{equation}
\alpha _{2}(t)=-\overline{a}_{1}(t)\frac{d_{12}}{d_{p}}\sum\limits_{k\neq n}%
\frac{kJ_{k}\left( Z\right) e^{-i(k-n)\omega _{0}t}}{k-n}.  \label{rap2}
\end{equation}%
It should be noted that the nonperturbative resonant approach used here put
the following restrictions:%
\begin{equation}
\left\vert F_{n}\right\vert ,\left\vert \Delta _{n}\right\vert ,\left\vert
\delta _{n}\right\vert \ll \omega _{0}  \label{applic}
\end{equation}%
on the characteristic parameters of the considered problem (for more details
see \cite{AM}). As it was mentioned in Introduction, this condition is not
fulfilled in Ref. \cite{GO}.

\subsection{Harmonic generation}

We restrict our study to the coherent part of the spectrum, i.e., the
spectrum of the mean dipole moment which is predominant for the forward
direction when the number of scatterers is relatively large \cite{EF,LFE}.
In the Schr\"{o}dinger picture the coherent part of the spectrum is \cite{EF}%
:%
\begin{equation}
S(\omega )=\left\vert \int_{-\infty }^{\infty }dte^{-i\omega t}\left\langle
D(t)\right\rangle \right\vert ^{2},  \label{Schrod_pic}
\end{equation}%
where%
\begin{equation}
\left\langle D(t)\right\rangle =\left\langle \Psi (t)\right\vert \mathbf{%
\hat{e}}\cdot \mathbf{\hat{d}(0)}\left\vert \Psi (t)\right\rangle ,
\label{dev1}
\end{equation}%
is the time-dependent mean dipole moment, which is the main observable
quantity. With the help of the wave function (\ref{wf}) the expectation
value of the dipole operator (\ref{dev1}) can be written as:%
\begin{equation*}
\left\langle D(t)\right\rangle =d_{11}\left\vert a_{1}(t)\right\vert
^{2}+d_{22}\left\vert a_{2}(t)\right\vert ^{2}
\end{equation*}%
\begin{equation}
+\left\{ d_{12}a_{1}^{\ast }a_{2}\sum\limits_{k}J_{k}\left( Z\right)
e^{i(k-n)\omega _{0}t}+\mathrm{c.c.}\right\} .  \label{dev2}
\end{equation}%
where $a_{1,2}(t)=\overline{a}_{1,2}(t)+\alpha _{1,2}(t)$. Combining Eqs. (%
\ref{diff_sol}), (\ref{rap1}), (\ref{rap2}), and (\ref{dev2}) one can
calculate analytically the expectation value of the dipole operator for an
arbitrary initial atomic state. The Fourier transform of $\left\langle
D(t)\right\rangle $ gives the coherent part of the dipole spectrum.

The solution (\ref{diff_sol}) for the system initially situated in the
ground state, is:%
\begin{equation}
\overline{a}_{1}(t)=e^{i\delta _{n}t/2}\left[ \cos \left( \frac{\Omega _{n}t%
}{2}\right) -i\frac{2\Delta _{n}+\delta _{n}}{\Omega _{n}}\sin \left( \frac{%
\Omega _{n}t}{2}\right) \right] ,  \label{avg1}
\end{equation}%
\begin{equation}
\overline{a}_{2}(t)=i\frac{2F_{n}}{\Omega _{n}}e^{i\delta _{n}t/2}\sin
\left( \frac{\Omega _{n}t}{2}\right) ,  \label{avg2}
\end{equation}%
which expresses the Rabi oscillations with frequency 
\begin{equation}
\Omega _{n}\equiv \sqrt{4F_{n}^{2}+(2\Delta _{n}+\delta _{n})^{2}}.
\label{Rabi1}
\end{equation}%
The generalized Rabi frequency at $n$-photon resonance has nonlinear
dependence on the amplitude of a pump wave field. As we can see from (\ref%
{avg1}), (\ref{avg2}) the amplitudes of Rabi oscillations are diminished due
to dynamic Stark effect. The destructive effect of the latter can be
compensated by an appropriate detuning: 
\begin{equation}
\delta _{n}=-2\Delta _{n}.  \label{des detuning}
\end{equation}%
Replacing the probability amplitudes in (\ref{dev2}) by the corresponding
expressions (\ref{avg1}) and (\ref{avg2}), one can derive the final
analytical expression for $\left\langle D(t)\right\rangle $. As is seen from
Eqs. (\ref{avr}), (\ref{stark}), the multiphoton coupling is proportional to
the ratio $d_{12}/d_{p}$, while the dynamic Stark shift is proportional to $%
d_{12}^{2}/d_{p}^{2}$. Since large dynamic Stark shifts make difficult the
maintenance of considerable population transfer, here we consider systems
with $\mu =\left\vert d_{12}\right\vert /d_{p}<<1$. Taking into account the
smallness of the parameter $\mu $, from (\ref{dev2}) in the first order of
approximation by the small parameter $\mu $ we obtain the following compact
analytic formula:%
\begin{equation}
\left\langle D(t)\right\rangle =C_{0}+\sum\limits_{k\neq 0}[S_{k}\sin
(k\omega _{0}t)+C_{k}\cos (k\omega _{0}t)],  \label{main}
\end{equation}%
where%
\begin{equation*}
C_{0}=d_{11}+\left( d_{p}\frac{2F_{n}^{2}}{\Omega _{n}^{2}}-d_{12}\frac{%
2F_{n}}{\Omega _{n}}\frac{2\Delta _{n}+\delta _{n}}{\Omega _{n}}J_{n}\left(
Z\right) \right)
\end{equation*}%
\begin{equation}
\times (1-\cos (\Omega _{n}t)),  \label{THz}
\end{equation}%
describes the low-frequency part of the spectrum, and%
\begin{equation}
S_{k}=d_{12}\frac{2F_{n}}{\Omega _{n}}\frac{nJ_{n+k}\left( Z\right) }{k}\sin
(\Omega _{n}t),  \label{sin}
\end{equation}%
\begin{equation}
C_{k}=d_{12}\frac{2F_{n}}{\Omega _{n}}\frac{2\Delta _{n}+\delta _{n}}{\Omega
_{n}}\frac{nJ_{n+k}\left( Z\right) }{k}(1-\cos (\Omega _{n}t)),  \label{cos}
\end{equation}%
describe harmonic radiation\textrm{. }As we can see from (\ref{sin})-(\ref%
{cos}), the harmonic spectrum consists of triplets, with harmonic and two
hyper-Raman lines with frequencies displaced by $\Omega _{n}$. From (\ref%
{Rabi1}) follows that frequencies of hyper-Raman and low-frequency lines can
be smoothly tuned via variation of the laser field strength and detuning.

The expression for $\left\langle D(t)\right\rangle $ shows that intensities
of the harmonics mainly determined by the behaviour of Bessel function.\
Since the Bessel function $J_{m}\left( Z\right) $ steeply decreases with
increase of index $m\gtrsim Z$, then the cut-off harmonic $s_{c}$ is
determined from the condition $s_{c}-n\sim Z$. Then, from the estimation for
cut-off harmonic follows that upper limit of the frequency which can be
effectively generated by direct $n$-photon excitation, is higher for the
systems with larger difference of energy and dipole moments in stationary
states ($\omega _{c}\sim \varepsilon _{2}-\varepsilon _{1}+d_{p}E_{0}$)%
\textrm{\ }. Although cut-off frequency is not directly depend on the photon
order $n$\textrm{\ }of the resonance, but taking into account harmful
ionization, the large $n$ is preferred for efficient harmonic generation at
the given strength of the laser field. Note that due to the oscillating
character of Bessel function, harmonics can be strengthened or weakened via
variation of the laser field strength within the resonance width. 
\begin{figure}[tbp]
\includegraphics [width=83mm,height=60mm] {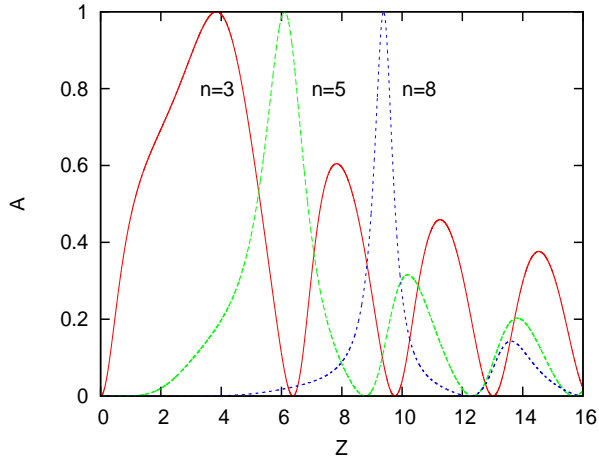}
\caption{(Color online) The dependence of the amplitude of state-populations
oscillation $A=4F_{n}^{2}/\Omega _{n}^{2}$ on parameter $Z$ for zero
detuning ($\protect\delta _{n}=0$) and $d_{12}/d_{p}=0.1$. The solid (red)
line corresponds to the three-photon ($n=3$) resonance; the dashed (green)
line corresponds to the five-photon ($n=5$) resonance; the dotted (blue)
line corresponds to the eight-photon ($n=8$) resonance. }
\end{figure}

The intensity of the harmonic components is proportional to $%
F_{n}^{2}/\Omega _{n}^{2}$, which is the amplitude of population
oscillations of the states. Thus, for the effective radiation generation one
should provide a considerable population transfer between the ground and
excited states, i.e., $\left\vert F_{n}\right\vert \sim \Omega _{n}$. As it
follows from (\ref{avr}), (\ref{stark}), and (\ref{Rabi1}), besides the
ratio $d_{12}/d_{p}$, parameter $Z$ may also have substantial effect on
population transfer. In Fig. 1 we plot dependence of the amplitude of
population oscillations of the states on parameter $Z$, for zero detuning
and different values of the photon order $n$. The complete population
transfer is achieved when\textrm{\ }$\Delta _{n}=0$\textrm{. }In fact, for
considerable population transfer it is important to compensate dynamic Stark
shift by an appropriate detuning, specifically for high-order multiphoton
excitation in the weaker laser fields.

The pattern of dipole intensities in harmonic triplet is determined from the
expressions (\ref{sin})-(\ref{cos}): 
\begin{equation}
\frac{I_{k\omega _{0}\pm \Omega _{n}}}{I_{k\omega _{0}}}=\frac{1}{4}\left[ 
\frac{J_{n-k}\left( Z\right) +J_{n+k}\left( Z\right) }{J_{n-k}\left(
Z\right) -J_{n+k}\left( Z\right) }\frac{\Omega _{n}}{2\Delta _{n}+\delta _{n}%
}\mp 1\right] ^{2}.  \label{compare}
\end{equation}%
In general, due to oscillating character of the Bessel functions, the
triplet pattern can be significantly changed depending on the pump wave
field strength. In moderately strong laser fields $Z\lesssim 1$ and in case
of considerable population transfer, i.e., when $\Omega _{n}\simeq
\left\vert 2\Delta _{n}+\delta _{n}\right\vert $, one of the hyper-Raman
lines becomes weaker than the other two lines of the triplet. Distribution
of intensities is essentially changed for almost complete population
transfer, i.e., when $\Omega _{n}\approx 2\left\vert F_{n}\right\vert \gg
\left\vert 2\Delta _{n}+\delta _{n}\right\vert $. In this case the
intensities of the hyper-Raman components become on the same order and
harmonic component practically vanishes. The latter is proportional to $\mu
^{2}$ and appears only in the next order of approximation by the small
parameter $\mu $.

\subsection{LOW-FREQUENCY RADIATION}

As it has been mentioned above, the expectation value of the dipole operator
(\ref{main}) also has low-frequency term $\left\langle D_{\mathrm{low}%
}(t)\right\rangle $ which is given by the formula:%
\begin{equation*}
\left\langle D_{\mathrm{low}}(t)\right\rangle =-d_{p}\left[ \frac{2F_{n}^{2}%
}{\Omega _{n}^{2}}-\frac{d_{12}}{d_{p}}\frac{2F_{n}}{\Omega _{n}}\right.
\end{equation*}%
\begin{equation}
\left. \times \frac{2\Delta _{n}+\delta _{n}}{\Omega _{n}}J_{n}\left(
Z\right) \right] \cos \left( \Omega _{n}t\right) ,  \label{Thzd}
\end{equation}%
according to Eq. (\ref{THz}). Formula (\ref{Thzd}) describes the radiation
at the smoothly tunable frequency $\Omega _{n}$, i.e., at the frequency of
oscillations of the population inversion. This low-frequency radiation also
depends on the amplitude of population oscillations of the states. According
to Eq. (\ref{Thzd}), with increase of the amplitude of Rabi oscillations the
intensity of low-frequency radiation increases. When population transfer is
almost complete, e.g., when dynamic Stark shift is compensated by
appropriate detuning, the low-frequency term can be written as:%
\begin{equation}
\left\langle D_{\mathrm{low}}(t)\right\rangle =-\frac{d_{p}}{2}\cos (\Omega
_{R}t),  \label{THzSt}
\end{equation}%
where the Rabi frequency $\Omega _{R}$ is determined by the expression: 
\begin{equation}
\Omega _{R}=\left\vert 2\frac{d_{12}}{d_{p}}n\omega _{0}J_{n}\left( Z\right)
\right\vert .  \label{Rabi}
\end{equation}%
In this case the emitted radiation intensity does not depend on the pump
wave field strength and is determined only by the difference of PDM in the
excited and ground states.

For the considered problem, the Rabi frequency may be varied within the
following limits:\textrm{\ }$2\pi /T_{r}<<\Omega _{R}<<\omega _{0}$, where%
\textrm{\ }$T_{r}$\textrm{\ }describes the relaxation time in the molecular
system. Hence, for electronic transitions\textrm{\ }$\Omega _{R}$ can lie
from microwave to infrared frequencies. Here, we concentrate on terahertz
region of electromagnetic spectrum ($0.1-10$ $\mathrm{THz}$) due to the
important properties of such radiation and its numerous applications for
spectroscopy, sensing, imaging, and etc (for a review see \cite{THzRev}).

For comparison of considered scheme with the other schemes of THz radiation,
we make some estimations for total radiation power of the ensemble of $N$
molecular emitters. It is assumed that the molecules are oriented in the
same direction. Though, in case of freely rotating molecules
temperature-dependent distribution of molecular orientations reduces
transition and permanent dipole moments, the intensity of emitted radiation
also will be reduced to some extent. Due to phase-matching factor the
coherent radiation entirely occurs almost in forward direction. The power of
low-frequency radiation was evaluated for resonantly driven three-level
atomic-molecular ensemble in \cite{THz} and will not be repeated here in
detail. We will adopt the estimations made for circular cylinder model of
superradiant emitter (in a limit of long needle) for ensemble of 2-level
polar molecules. The radiation power for ensemble of polar molecules
coherently excited by Gaussian laser beam with a waist $w_{0}$ can be
approximated by formula:%
\begin{equation}
P\simeq \frac{d_{p}^{2}\Omega _{R}^{3}}{16c^{2}}\frac{\pi ^{3}w_{0}^{4}}{%
n^{2}}\Delta _{\Omega }^{2}L_{c}N_{0}^{2},  \label{pow}
\end{equation}%
where $\Delta _{\Omega }$ is a degree of monochromaticity specified by
driven field variation in the interaction region, $L_{c}$ is the coherent
length over which the low-frequency can be built up coherently and $N_{0}$
is the density of polar emitters. The corresponding incident pulse duration
should be: $\tau \succsim 2\pi n/(\Omega _{R}\Delta _{\Omega })$. As we can
see from (\ref{pow}), it is more desirable to generate THz radiation via
low-order multiphoton excitation. Here we make estimations for two-photon
resonance ($n=2$); the incident laser beam waist is taken to be: $%
w_{0}\succsim 1\ \mathrm{mm}$, the coherence length: $L_{c}\succsim 10\ 
\mathrm{cm}$, the difference of PDM in the excited and ground states: $%
d_{p}\sim 5\ \mathrm{a.u.}$ and emitters density: $N_{0}\sim 10^{15}\ 
\mathrm{cm}^{-3}$. For the moderate monochromaticity $\Delta _{\Omega
}=\delta \Omega _{R}/\Omega _{R}\sim 0.1$ at $\Omega _{R}/(2\pi )\simeq 3\ 
\mathrm{THz}$, the incident pulse duration should be: $\tau \succsim 6\ 
\mathrm{ps}$. Estimated total radiation power $P\simeq 100\ \mathrm{W}$ is
comparable with parameters for most powerful THz sources. Note that
submillimeter-sized array of resonantly driven 2-level quantum dots with
induced dipole moment $\sim 10\ \mathrm{D}$ gives the power of microwatt
level \cite{Kib}.

In case of a solid medium, the radiation power could be larger due to higher
densities of matter. The other advantage of the solid medium is the degree
of anisotropy, since molecules can practically be oriented in the same
direction, e.g., in some organic crystals or when the molecules are inserted
in a solid matrix that is transparent to the radiation. The volume of
coherent radiation for a solid medium is $\sim \lambda ^{3}$, and the number
of polar emitters can be estimated by $N\sim \lambda ^{3}N_{0}$, which at
densities $N_{0}\sim 10^{21}\mathrm{cm}^{-3}$ gives $N$ $\sim 10^{15}$. The
rough estimations show that using submillimeter-sized crystals with $\sim $%
1\% ordered dipole moments of emitters one can achieve giant total radiation
power for coherent THz radiation up to megawatt level.

\section{NUMERICAL RESULTS AND\ DISCUSSION}

In this section we present numerical solutions of\ time-dependent Schr\"{o}%
dinger equation for a two-level model with Hamiltonian (\ref{Ham}). The set
of equations for the probability amplitudes has been solved using a standard
fourth-order Runge-Kutta algorithm \cite{Num}. The Fourier transformations
for estimation of power spectra are performed using the fast Fourier
transform technique. For smooth turn-on of pump wave field, the latter is
described by the envelope in hyperbolic tangent $\tanh (t/\tau )$ form,
where $\tau $ characterizes the turn-on time and chosen to be $20\pi /\omega
_{0}$. The transition frequency is set to be $\varepsilon _{2}-\varepsilon
_{1}=0.2$ \textrm{a.u.} and lies in UV region, but the results can be scaled
to different energy regimes. The transition dipole moment is chosen to be%
\textrm{\ }$d_{12}=0.5$\textrm{\ a.u., }and difference of PDM\textrm{\ }in
the excited and ground states: $d_{p}=5$\textrm{\ a.u. }(except where it is
declared otherwise). It is assumed that initially all the population is in
the ground state $\left\vert 1\right\rangle .$

Figure 2 shows dipole spectrum (coherent part) as a function of harmonic
order for four-photon ($n=4$) resonant excitation. The pump field strength
is $E=0.01$ \textrm{a.u. }and multiphoton detuning $\delta _{n}$ is set to
zero. The solid (red) line corresponds to numerical calculations, while the
dashed (green) line corresponds to the approximate expression (\ref{main}).
For better visibility, the spectrum corresponding to analytical calculations
has been slightly shifted to the right. As we can see from the Fig. 2, the
analytical formula (\ref{main}) is in good agreement with the exact results.
For employed field strengths the triplet structure of harmonics and
low-frequency line are not clearly seen, hence the latter is illustrated in
the inset.

\begin{figure}[tbp]
\includegraphics [width=83mm,height=60mm] {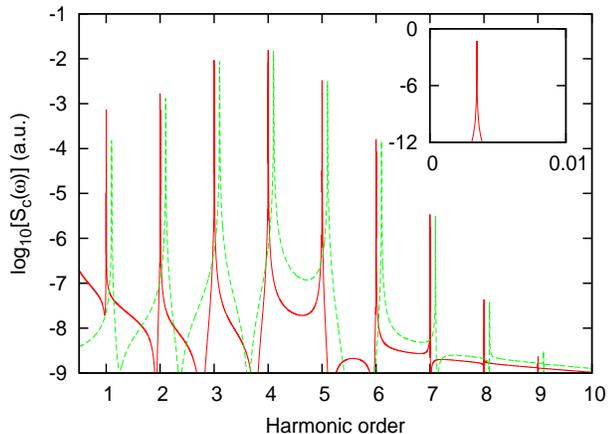}
\caption{(Color online) The logarithm of the coherent part of the spectrum S$%
_{C}(\protect\omega )$ at four-photon ($n=4$) resonance with zero detuning ($%
\protect\delta _{n}=0$). The energy level difference is $\protect\varepsilon %
_{2}-\protect\varepsilon _{1}=0.2$ $\mathrm{a.u.}$, laser field strength $%
E=0.01$ $\mathrm{a.u.}$, and dipole moments are $d_{12}=0.5$ $\mathrm{a.u.}$
and $d_{p}=5$ $\mathrm{a.u.}$. The solid (red) line corresponds to numerical
calculations; the dashed (green) line corresponds to the approximate
solution (for better visibility the latter has been slightly shifted to the
right). The inset shows the low-frequency peak at Rabi frequency.}
\end{figure}

\begin{figure}[tbp]
\includegraphics [width=83mm,height=60mm] {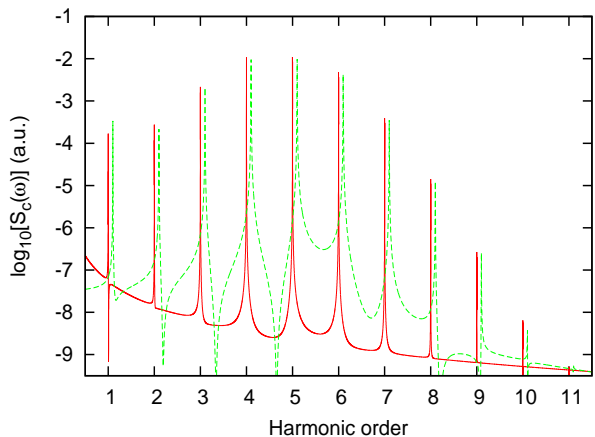}
\caption{(Color online) The logarithm of the coherent part of the spectrum S$%
_{C}(\protect\omega )$ at five-photon ($n=5$) resonance. The energy level
difference is $\protect\varepsilon _{2}-\protect\varepsilon _{1}=0.2$ $%
\mathrm{a.u.}$, detuning $\protect\delta _{n}=0.00014$ $\mathrm{a.u.,}$
laser field strength $E=0.01$ $\mathrm{a.u.}$, and dipole moments are $%
d_{12}=0.5$ $\mathrm{a.u.}$ and $d_{p}=5$ $\mathrm{a.u.}$. The solid (red)
line corresponds to numerical calculations; the dashed (green) line
corresponds to the approximate solution (for better visibility the latter
has been slightly shifted to the right). }
\end{figure}

Dipole spectrum as a function of harmonic order for five-photon ($n=5$)
resonant excitation with almost complete population transfer is plotted in
Fig. 3. The pump field strength is $E=0.01$ \textrm{a.u.} and detuning for
compensation of dynamic Stark shift is $\delta _{n}=0.00014$ \textrm{a.u.}.
As in Fig 2, the solid (red) line corresponds to numerical calculations and
the dashed (green) line corresponds to the approximate expression (\ref{main}%
). The numerical results confirm the analytical expression for expectation
value of the mean dipole moment. These figures confirm the estimation $%
s_{c}-n\sim Z$ for cut-off position.\textbf{\ }

In Fig. 4 we plot characteristic harmonic triplet (here, for 6th harmonic)
under the same conditions of excitation, as in Fig. 3. The solid (red) line
corresponds to detuning, compensating dynamic Stark shift, and dashed
(green) line corresponds to zero detuning. For better visibility, the
spectrum corresponding to compensating detuning has been slightly shifted to
the right. The numerical results confirm estimations for lines position in
harmonic triplet and the fact that via compensation of dynamic Stark shifts
one can attain the higher intensities. Distribution of intensities,
particularly, damping of harmonic lines in case of compensating detuning and
lower-frequency hyper-Raman lines at zero detuning, have been predicted in
section II.

To demonstrate the effects of PDM we plot the spectrum of the two-level
system without PDM under the same conditions of excitation, as in Fig. 3. As
is shown in Fig. 5, there are fewer harmonics and with negligibly small
amplitudes. Practically the full radiation is concentrated on the incident
radiation frequency. Even harmonics, as well as the low-frequency radiation
are absent because of inversion symmetry. Note that these essential
distinctions in intensities of the harmonics are for the pump field
strengths with negligible ionization and dynamic Stark shifts. To achieve
the efficient harmonic generation when $d_{p}=0$, required laser fields
should be comparable to characteristic fields of considered system: $%
E_{0}\gtrsim \left( \varepsilon _{2}-\varepsilon _{1}\right) /d_{12}$, which
will cause complete ionization.

\begin{figure}[tbp]
\includegraphics [width=83mm,height=60mm] {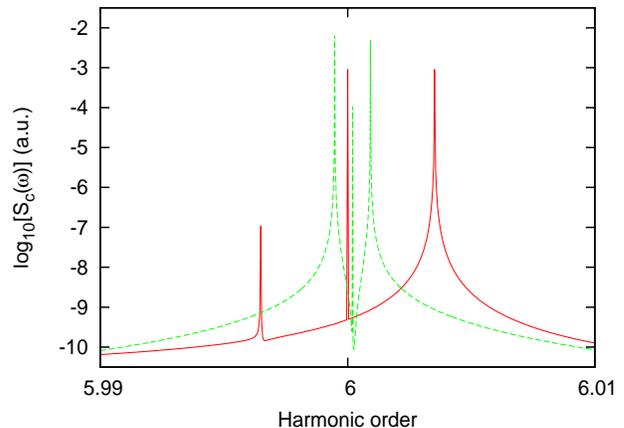}
\caption{(Color online) The structure of satellite peaks around the 6th
harmonics under the same conditions of excitation as in Fig. 2. The solid
(red) line corresponds to zero detuning ($\protect\delta _{n}=0$); the
dashed (green) line corresponds to detuning $\protect\delta _{n}=0.00014\ 
\mathrm{a.u.}$ (for better visibility the latter has been slightly shifted
to the right). }
\end{figure}

\begin{figure}[tbp]
\includegraphics [width=83mm,height=60mm] {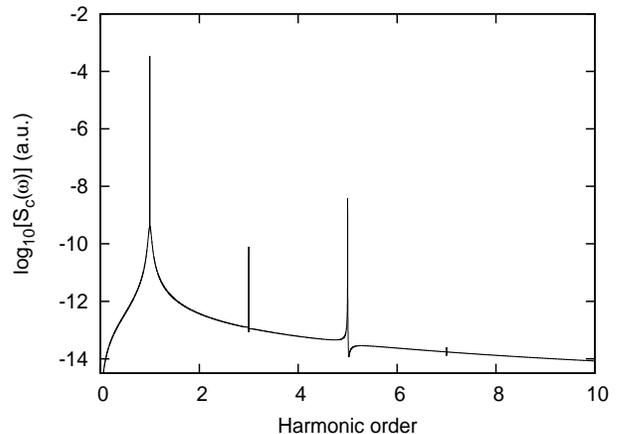}
\caption{Same as in figure 2, but for the case of zero dipole moments in
stationary states.}
\end{figure}

\begin{figure}[t]
\includegraphics [width=83mm,height=60mm] {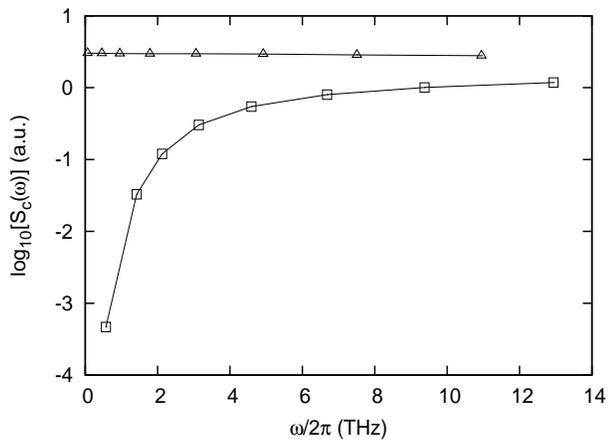}
\caption{The low-frequency emission rate as a function of the emission
frequency under the same conditions of excitation as in Fig. 2. (a) upper
line corresponds to compensated dynamic Stark shift. (b) lower line
corresponds to zero detuning ($\protect\delta _{n}=0$). }
\end{figure}

We also made numerical calculations for low-frequency part of the radiation
spectrum. Figure 6 displays the dependence of the emission rate on emission
frequency at five-photon resonance. The lower line corresponds to zero
detuning and upper line to detuning compensating dynamic Stark shift. The
calculations have been made for the same field strengths. As we can see from
the Fig. 6, the frequency and intensity of the emission lines appreciably
depend on the magnitude of the detuning and compensation of the Stark shift
is of considerable importance. The numerical results for spectral
characteristics are in good agreement with estimations based on (\ref{Thzd})
and (\ref{THzSt}).

It should be noted that evenly charged molecular ions at large internuclear
distances \cite{Gib}, driven flux qubits \cite{Oliver}, and hydrogenlike
atoms \cite{AM,AM2,AM3} are good examples of efficient direct multiphoton
excitation. The mentioned three-level configurations are also good
candidates for moderately high harmonics generation \cite{AAM}. However, in
contrast to 2-level model with PDM, due to inversion symmetry the spectrum
of three-level system does not contain even harmonics (with corresponding
hyper-Raman lines) and low-frequency components under the conditions of
excitation considered here.

\section{CONCLUSION}

We have presented a theoretical treatment of the dipole spectrum of a
two-level system possessing PDM under direct multiphoton resonant
excitation. With the help of nonperturbative resonant approach, we have
studied approximate stationary solution of the Schrodinger equation and
obtain a simple analytical expression for scattered coherent radiation. We
have also numerically investigated the harmonic spectra and effects of
compensation of the dynamic Stark shift on emission spectrum. As has been
shown, the compensation of dynamic Stark shift by appropriate detuning could
substantially improve efficiency of radiation generation. The numerical
results are in good agreement with obtained analytical results for
expectation value of dipole operator. We have also shown that the existence
of PDM enables the considerable population transfer in moderately strong
laser fields, and therefore could contribute to generation of higher
harmonics with sufficiently large amplitudes. In addition\textrm{, }due to
broken inversion symmetry of systems with PDM there is substantial emission
at frequency much smaller than the laser frequency. The considered scheme
may serve as a promising method for efficient generation of moderately high
harmonics and intense widely tunable low-frequency radiation. Note that for
coherent radiation of an molecular ensemble one needs more rigorous
treatment, which should account for degree of anisotropy in the\textrm{\ }%
orientational distribution of polar molecules, as well as collective effects
in the medium. Work in this direction is in progress and will be presented
in forthcoming paper.

\begin{acknowledgments}
This work was supported by SCS of RA under Project No. 10-3E-17 and
CRDF/NFSAT/SCS ECSP-09-72.
\end{acknowledgments}


\begin{thebibliography}{99}
\bibitem{HHG1} B. W. Shore and P. L. Knight, J. Phys. B \textbf{20}, 413
(1987); P. B. Corkum, Phys. Rev. Lett. \textbf{71}, 1994 (1993); M.
Lewenstein, Ph. Balcou, M. Yu. Ivanov, A. L'Huillier, and P.B. Corkum, Phys.
Rev. A\textbf{\ 49}, 2117 (1994).

\bibitem{HHG3} M. Protopapas, C. H. Keitel, and P. L. Knight, Rep. Prog.
Phys. \textbf{60}, 389 (1997); P. Sali\'{e}res A. L'Huillier, P. Antoine,
and M. Lewenstein, Adv. At., Mol., Opt. Phys. \textbf{41}, 83 (1999); T.
Brabec and F. Krausz, Rev. Mod. Phys. \textbf{72}, 545 (2000); P. Agostini
and L. F. DiMauro, Rep. Prog. Phys. \textbf{67}, 813 (2004).

\bibitem{QuantWells} P. Haljan, T. Fortier, P. Hawrylak, P. B. Corkum, and
M. Yu. Ivanov, Laser Phys. \textbf{13}, 452 (2003); D. Golde, T. Meier, and
S.W. Koch, Phys. Rev. B \textbf{77}, 075330 (2008).

\bibitem{LargeMol} Z. H. Kafafi, J. R. Linde, R. G. S. Pong, F. J. Bartoli,
L. J. Lingg, and J. Milliken, Chem. Phys. Lett. \textbf{188}, 492 (1992);
G.P. Zhang, Phys. Rev. Lett. \textbf{95}, 047401 (2005).

\bibitem{BoundCont} V. Averbukh, O. E. Alon, and N. Moiseyev, Phys. Rev. A 
\textbf{64}, 033411 (2001); C. F. deMorissonFaria and I. Rotter, Phys. Rev.
A \textbf{66}, 013402 (2002).

\bibitem{AAM} H. K. Avetissian, B. R. Avchyan, and G. F. Mkrtchian, Phys.
Rev. A\textbf{\ 77}, 023409 (2008).

\bibitem{KrMi} V. P. Krainov and Z. S. Milyukov, Laser Phys.\textbf{\ 4},
544 (1994).

\bibitem{Duvall} R. E. Duvall, E. J. Valeo, and C. R. Oberman, Phys. Rev. A%
\textbf{\ 37}, 4685 (1988).

\bibitem{PDM} M. A. Kmetic and W. J. Meath, Physics Letters A \textbf{108},
340 (1984); A. Brown, W. J. Meath, and P. Tran, Phys. Rev. A \textbf{63},
013403 (2000); \textit{ibid}. \textbf{65}, 063401 (2002).

\bibitem{AM} H. K. Avetissian and G. F. Mkrtchian, Phys. Rev. A\textbf{\ 66}%
, 033403 (2002).

\bibitem{Gib} G. N. Gibson, Phys. Rev. Lett. \textbf{89}, 263001 (2002).

\bibitem{AM2} H. K. Avetissian, G. F. Mkrtchian, M. G. Poghosyan, Phys. Rev.
A \textbf{73}, 063413 (2006).

\bibitem{AM3} H. K. Avetissian, B.R. Avchyan, G. F. Mkrtchian, Phys. Rev. A 
\textbf{74}, 063413 (2006).

\bibitem{ABM} H. K. Avetissian, A. Brown, G. F. Mkrtchian, Phys. Rev. A 
\textbf{80}, 033413 (2009).

\bibitem{Kib} O. V. Kibis, G. Y. Slepyan, S. A. Maksimenko, and A. Hoffmann,
Phys. Rev. Lett. \textbf{102}, 023601 (2009).

\bibitem{THz} H. K. Avetissian, B. R. Avchyan, and G. F. Mkrtchian, Phys.
Rev. A\textbf{\ 82}, 063412 (2010).

\bibitem{Calderon} O. G. Calder\'{o}n, R. Guti\'{e}rrez-Castrej\'{o}n, and
J. M. Guerra, IEEE J. Quantum Electron. \textbf{35}, 47 (1999).

\bibitem{GO} V. P. Gavrilenko and E. Oks, J. Phys. B\textbf{\ 33}, 1629
(2000).

\bibitem{P1} D. L. Andrews and W. J. Meath, J. Phys. B\textbf{\ 26}, 4633
(1993).

\bibitem{P2} A. Brown and W. J. Meath, Phys. Rev. A\textbf{\ 53}, 2571
(1996).

\bibitem{P3} S. H. Nilar, A. J. Thakkar, A. E. Kondo, and W. J. Meath, Can.
J. Chem. \textbf{71}, 1663 (1993).

\bibitem{P4} V. A. Kovarskii, Phys. Usp. \textbf{42}, 797 (1999).

\bibitem{TC-B} D. A. Telnov and Shih-I Chu, Phys. Rev. A\textbf{\ 71},
013408 (2005); A. D. Bandrauk, S. Chelkowski, D. J. Diestler, J. Manz, and
K.-J. Yuan, Phys. Rev. A\textbf{\ 79}, 023403 (2009).

\bibitem{Watson} G. N. Watson, \textit{A Treatise on the Theory of Bessel
Functions} (Cambridge University Press, Cambridge, 1944).

\bibitem{EF} J. H. Eberly and M. V. Fedorov, Phys. Rev. A\textbf{\ 45}, 4706
(1992).

\bibitem{LFE} D. G. Lappas, M. V. Fedorov, and J. H. Eberly, Phys. Rev. A%
\textbf{\ 47}, 1327 (1993).

\bibitem{THzRev} P. H. Siegel, IEEE Trans. Microw. Theory Tech., \textbf{50}%
, 910 (2002); D. Mittleman, \textit{Sensing with Terahertz Radiation},%
\textit{\ }(Springer, Heidelberg, 2003); D. Dragoman and M. Dragoman,
Progress in Quantum Electronics \textbf{28}, 1 (2004); D. W. Woolard, E. R.
Brown, M. Pepper, and M. Kemp, Proc. IEEE \textbf{93}, 1722 (2005); M.
Tonouchi, Nature Photonics \textbf{1}, 97 (2007).

\bibitem{Num} W. H. Press, S. A. Teukolsky, W. T. Vetterling, and B. P.
Flannery, \textit{Numerical Recipes in C} ( Cambridge University Press,
Cambridge, 1992).

\bibitem{Oliver} W. D. Oliver, Y. Yu, J. C. Lee, K. K. Berggren, L. S.
Levitov, and T. P. Orlando, Science \textbf{310}, 1653 (2005).
\end{thebibliography}
\end{document}